\documentclass[aps,twocolumn,prl]{revtex4}
\usepackage{amsmath,amssymb,bm}
\usepackage{graphicx}
\usepackage{epstopdf}
\usepackage{latexsym}

\begin{document}
\title{Dynamics and transport of the $Z_2$ spin liquid:\\ application to $\kappa$-(ET)$_2$Cu$_2$(CN)$_3$}
\author{Yang Qi}
\affiliation{Department of Physics, Harvard University, Cambridge MA 02138, USA}
\author{Cenke Xu}
\affiliation{Department of Physics, Harvard University, Cambridge MA 02138, USA}
\author{Subir Sachdev}
\affiliation{Department of Physics, Harvard University, Cambridge MA 02138, USA}
\date{September 2, 2008}
\begin{abstract}
We describe neutron scattering, NMR relaxation, and thermal transport properties of
$Z_2$ spin liquids in two dimensions.
Comparison to recent experiments on the spin $S=1/2$ triangular lattice antiferromagnet in
$\kappa$-(ET)$_2$Cu$_2$(CN)$_3$ shows that this compound may realize a $Z_2$ spin
liquid. We argue that the topological `vison' excitations dominate thermal transport, and that recent
thermal
conductivity experiments by M. Yamashita {\em et al.} have observed the vison gap.
\end{abstract}

\maketitle

Much attention \cite{kanoda0,kanoda1,kanoda2,yamashita,mot,palee,qi} has recently
focused on the organic compound $\kappa$-(ET)$_2$Cu$_2$(CN)$_3$
because it may be the first experimental realization of a resonating valence bond spin liquid \cite{fazekas,krs}.
This compound belongs to a class \cite{powell,kato6} of organic Mott insulators 
which can be described by $S=1/2$ spins residing on the vertices of a triangular lattice. Experiments
have not detected any magnetic order or a structural distortion leading to a doubling of the unit cell in $\kappa$-(ET)$_2$Cu$_2$(CN)$_3$,
and so there is justifiable optimism that the elusive spin liquid state may finally have been found.

The debate then turns to the identification of the precise spin
liquid state, among the many possible candidates. Measurements of
the electronic specific heat, $C_P$, by S.~Yamashita {\em et al.}
\cite{kanoda2} were interpreted to yield a non-zero low
temperature ($T$) value of $\gamma = \lim_{T \rightarrow 0} C_P
/T$. Such a non-zero $\gamma$ is characteristic of a Fermi
surface, and hence a spin liquid state with a Fermi surface of
neutral, $S=1/2$, fermionic spinons was postulated
\cite{kanoda2,mot,palee}. However, it should be noted that the
measurement of $\gamma$ involves a potentially dangerous
subtraction of a divergent nuclear specific heat \cite{kanoda2}.

Very recently, M.~Yamashita {\em et al.} have measured \cite{yamashita} the thermal conductivity, $\kappa$,
to below $T \approx 0.1$ K. This has the advantage of focusing on the mobile excitations, and not
being contaminated by a nuclear contribution. A spinon Fermi surface should yield a non-zero
low $T$ limit for $\kappa/T$, but this quantity was clearly observed to vanish. Instead, the measured
$\kappa$ was fit reasonably well by the activated
behavior $\kappa \sim \exp (- \Delta_\kappa /T)$, with a `gap' $\Delta_\kappa \approx
0.46$ K. Furthermore, $\kappa$ was found to be insensitive to an applied field for $H < 4$ T,
suggesting that the gap $\Delta_\kappa$ is associated with a spinless excitation.
These observations appear to be incompatible with spinon Fermi surface states at these low $T$, and
we shall present an alternative theory here.

Also of interest are the measurements \cite{kanoda1} of the NMR relaxation rate, $1/T_1$.
The power-law behavior $1/T_1 \sim T^{a}$, with the exponent $a \approx 1.5$, was observed
for $0.02 < T < 0.3$ K. This requires the presence of spinful excitations with a gapless spectrum at the
fields of the NMR experiment, although at zero field there may well be a small spin gap.

In this paper, we will compare these observations with the
$Z_2$ spin liquid state originally proposed in Refs.~\onlinecite{rs1,jalabert,sstri}. The low energy excitations of this
state are
described by a $Z_2$ gauge theory, and the spinful
excitations are constructed from $S=1/2$ quanta (the spinons) which carry a $Z_2$ electric
charge. Crucial to our purposes here are vortex-like spinless excitations \cite{rc} which carry $Z_2$
magnetic flux, later dubbed `visons' \cite{sf}. A number of solvable models of $Z_2$ spin liquids,
with spinon and vison excitations, have been constructed \cite{sf,sondhi,kitaev,wen,freedman,wang,misguich}.
We propose here that it is the visons which dominate
the thermal transport in $\kappa$-(ET)$_2$Cu$_2$(CN)$_3$, and
the gap $\Delta_\kappa$ is therefore identified with a vison energy gap, $\Delta_v$. If our interpretation is correct, the vison
has been observed by M.~Yamashita {\em et al.} \cite{yamashita}.

Our proposal requires
that the density of states of low energy vison excitations is much larger than that of all other excitations.
A model appropriate to $\kappa$-(ET)$_2$Cu$_2$(CN)$_3$ is the triangular lattice $S=1/2$ antiferromagnet
with nearest neighbor two-spin exchange ($J_2$) and plaquette four-spin ($J_4$) exchange which was studied
by Liming {\em et al.} \cite{liming}. They found antiferromagnetic order at $J_4=0$ (as in earlier work \cite{singh}),
and a quantum phase transition to a spin liquid state with a spin gap around $J_4/J_2 \approx 0.1$. Notably, they found a very
large density of low-lying spin singlet excitations near the transition. We propose here that $\kappa$-(ET)$_2$Cu$_2$(CN)$_3$
is near this quantum phase transition, and identify these singlets with visons which have a small gap
and bandwidth, both much smaller than the spin exchange $J_2 \sim 250$K.
We will argue below that at $ T\ll J_2$, and comparable to the vison bandwidth, visons will dominate the thermal transport.

Further support for the proximity of a magnetic ordering quantum critical point
comes from \cite{kato6} the closely related series of compounds X[Pd(dmit)$_2$]$_2$. By varying
the anisotropy of the triangular lattice by varying X, we obtain compounds with decreasing magnetic ordering critical temperatures,
until we eventually reach a compound with a spin gap and valence bond solid (VBS) order \cite{kato4}. 
In between is the compound \cite{kato7} with X=EtMe$_3$P
which has been proposed to be at the quantum critical point \cite{kato6}, and has properties similar to $\kappa$-(ET)$_2$Cu$_2$(CN)$_3$. Finally, series expansion studies \cite{wms} also place the triangular lattice antiferromagnet near a quantum critical point
between magnetically ordered and VBS states.

A description of the NMR experiments requires a theory for the spinon excitations of the
$Z_2$ spin liquid. The many models of $Z_2$ spin liquids \cite{rs1,jalabert,sstri,rc,sf,sondhi,kitaev,wen,freedman,wang,misguich}
have cases with either fermionic or bosonic spinons.
While
we do not find a satisfactory explanation for the NMR with fermionic spinons,
we show that a model \cite{rs1,jalabert,sstri} of bosonic spinons in a spin liquid close to  the quantum phase transition
to the antiferromagnetically ordered state (as found in the model of Liming {\em et al.} \cite{liming})
does naturally explain the $T$ dependence of $1/T_1$.
We shall show below that the quantum critical region for this transition leads to
$1/T_1 \sim T^{\bar{\eta}}$ with the exponent \cite{vicari,kim} $\bar{\eta} = 1.37$, reasonably close to
the measured value $a=1.5$. It is important to note that the vison gap, $\Delta_v$, remains non-zero across
this magnetic ordering critical point \footnote{In the ordered state, the visons have a logarithmic interaction,
and the self-energy of an isolated vison diverges logarithmically with system size}. Consequently, our interpretation
of the experiments remains valid even if the system acquires a small antiferromagnetic moment, as may be the case
in the presence of the applied magnetic field present in the NMR measurements.

The remainder of the paper presents a number of computations of the physical properties of $Z_2$ spin liquids,
and uses them to elaborate on the experimental interpretation sketched above.

We begin with a theory \cite{css} of the spinon excitations near the quantum critical point between the
magnetically ordered state and the $Z_2$ spin liquid. Here the low energy spinons are $S=1/2$
complex bosons $z_\alpha$, with $\alpha = \uparrow, \downarrow$ a spin index, and the low energy
imaginary time action is
\begin{equation}
\mathcal{S} = \frac{1}{g} \int d^2r d\tau \left[ |\partial_\tau z_\alpha|^2 + c^2 |\nabla_r z_\alpha |^2 \right],
\end{equation}
where $(r, \tau)$ are spacetime co-ordinates, $g$ is a coupling which tunes the transition to the spin liquid
present for some $g>g_c$, and $c$ is a spin-wave velocity. We impose the local constraint $\sum_\alpha |z_\alpha|^2 = 1$
in lieu of a quartic self-interaction between the spinons.
This theory has an emergent O(4) global symmetry \cite{azaria,kim} (which becomes manifest when $z_\alpha$ is written in terms of its real and imaginary components).
This symmetry is an enlargement of the SU(2) spin rotation symmetry, and we will neglect the irrelevant terms which reduce
the symmetry to SU(2).

\paragraph{(i) Dynamic spin susceptibility.}
The dynamic spin correlations of $\mathcal{S}$ near the quantum
critical point can be computed by the $1/N$ expansion on the O($N$) model, which has
been described elsewhere \cite{csy}. With an eye towards possible
future neutron scattering measurements, we first describe the
dynamic spin susceptibility, $\chi (k, \omega)$ as a function of
momentum $k$ and real frequency $\omega$. Here the momentum $k$ is
measured as a deviation from the ordering wavevector, $Q$, of the
antiferromagnetically ordered state. At $g=g_c$ and $T=0$, this
has the quantum-critical form
\begin{equation}
\chi (k, \omega) = \frac{\mathcal{A}}{(c^2 k^2 - \omega^2 ) ^{1- \bar{\eta}/2}}, \label{chi}
\end{equation}
where the exponent $\bar{\eta}$ is related to the scaling
dimension of the composite spin operator $\sim z_\alpha
\sigma^y_{\alpha\gamma}\vec{\sigma}_{\gamma\beta} z_\beta$
($\vec{\sigma}$ are the Pauli matrices), and is known with high
precision from field-theoretic studies \cite{vicari} ($\bar{\eta}
= 1.374(12)$) and Monte Carlo simulations \cite{kim} ($\bar{\eta}
= 1.373(2)$). The overall amplitude $\mathcal{A}$ is
non-universal, but the same $\mathcal{A}$ will appear in a number
of results below. Integrating Eq.~(\ref{chi}) over all $k$, we
obtain the local susceptibility $\chi_L (\omega)$, which is also
often measured in scattering experiments, again at $g=g_c$ and
$T=0$
\begin{equation}
\mbox{Im}\,\chi_L (\omega) = \frac{\mathcal{A}\, \mbox{sgn}(\omega)}{4c^2} \frac{\sin (\pi \bar{\eta}/2)}{\pi \bar{\eta}/2}
|\omega|^{\bar{\eta}} . \label{chil}
\end{equation}

Let us now move into the spin liquid state, with $g>g_c$, where
the spinons have an energy gap $\Delta_z$. The critical results in
Eqs. (\ref{chi}) and (\ref{chil}) will apply for $|\omega| \gg
\Delta_z$, but for $|\omega| \sim 2 \Delta_z$, we will have
spectra characteristic of the creation of a pair of spinons (we
set $\hbar=1$, although it appears explicitly in a few expressions
below). Computing the pair creation amplitude of non-interacting
spinons, we obtain a step-discontinuity threshold at $\omega =
\sqrt{c^2 k^2 + 4 \Delta_z^2}$ (at $T=0$). However, the spinons do
have a repulsive interaction with each other, and this reduces the
phase space for spinon creation at low momentum, as described in the supplementary
material; the actual threshold behavior is:
\begin{equation}
\mbox{Im} \chi(k, \omega) = \frac{\mathcal{A}\, \mathcal{C}  \,
\mbox{sgn}(\omega)}{\Delta_z^{2-\bar{\eta}}} \, \frac{\theta
\left(|\omega| - \sqrt{k^2 + 4 \Delta_z^2} \right)}{\ln^2 \left(
\displaystyle \frac{ \left| \omega^2 - k^2 - 4 \Delta_z^2
\right|}{16 \Delta_z^2}\right) }, \label{chi2}
\end{equation}
where $\mathcal{C}$ is a universal constant; to leading order in the $1/N$ expansion,
 $\mathcal{C} = N^2/ 16$.
We can also integrate the $k$-dependent generalization of
Eq.~(\ref{chi2}) to obtain a threshold behavior for the local
susceptibility at $2 \Delta_z$: $ \mbox{Im}\,\chi_L (\omega) \sim
\mbox{sgn} (\omega) (|\omega| - 2 \Delta_z)/ \ln^2 (|\omega| - 2
\Delta_z) $.

\paragraph{(ii) NMR relaxation.} Turning to the NMR relaxation rate, we have to consider $T>0$, and compute
\begin{equation}
\Gamma = \lim_{\omega \rightarrow 0} \frac{k_B T}{\omega} \mbox{Im} \chi_L (\omega).
\end{equation}
This is far more subtle than the computations at $T=0$, because we
have to compute the damping of the quantum critical excitations at
$T>0$ and extend to the regime $\omega \ll T$. From general
scaling arguments \cite{csy}, we have
\begin{equation}
\Gamma = \frac{\mathcal{A}}{c^2} (k_B T)^{\bar{\eta}} \Phi (\Delta_z /(k_B T)), \label{Gamma}
\end{equation}
where $\Phi$ is a universal function. The computation of $\Phi$
for undamped spinons at $N=\infty$ is straightforward, and unlike
the case for confining antiferromagnets \cite{csy}, yields a
reasonable non-zero answer: $\Phi (y) = [4 \pi e^{y/2} (1 +
\sqrt{4 + e^y})]^{-1}$. However, the $1/N$ corrections are
singular, because $\Gamma$ has a singular dependence upon the
spinon lifetime. A self-consistent treatment of the spinon damping
is described in the supplementary material, and leads to 
the quantum-critical result ($\Delta_z = 0$):
\begin{equation}
\Phi (0) = \frac{(\sqrt{5}-1)}{16 \pi}\left(1  + 0.931 \frac{\ln N}{N} + \ldots \right).
\end{equation}

\paragraph{(iii) Thermal conductivity.}
We now turn to the thermal transport co-efficient measured in the
recent revealing experiments of Ref.~\onlinecite{yamashita}. We consider
the contribution of the spinons and visons in turn below, presenting further arguments
on why the vison contribution can dominate in the experiments.

\paragraph{(iii.a) Spinons.}
For agreement with the NMR measurements of $1/T_1$ \cite{kanoda1},
we need the spinons to be in the quantum critical regime, as
described above. Therefore, we limit our considerations here to
the quantum critical thermal conductivity of the spinons,
$\kappa_z$, with $\Delta_z = 0$. This can be obtained from the
recent general theory of quantum critical transport
\cite{hartnoll} which yields
\begin{equation}
\kappa_z = s c^2 \tau_z^{\rm imp},
\end{equation}
where $s$ is the entropy density of the spinons, and
$1/\tau_z^{\rm imp}$ is the spinon momentum relaxation rate, with the $T$
dependence  \begin{eqnarray} \tau_z^{\rm imp} \sim T^{2/\nu - 3}.
\label{e1} \end{eqnarray} Here $\nu$ is the critical exponent of the O(4)
model \cite{hasen}, $\nu = 0.749(2)$, and so $\tau_z^{\rm imp}
\sim T^{-0.33}$. The two dimensional entropy density can be
obtained from the results of Ref.~\onlinecite{csy}:
\begin{equation}
s = \frac{3 N \zeta (3) k_B^3 T^2}{2 \pi \hbar^2 c^2} \left[ \frac{4}{5}
- \frac{0.3344}{N} + \ldots \right],
\end{equation}
where $\zeta$ is the Reimann zeta function.
We estimate the co-efficient in Eq.~(\ref{e1}) in the supplementary material using
a soft-spin theory with the spinons moving in a random potential,
$V(r) |z_\alpha|^2$, due to impurities of density $n_{\rm imp}$
each exerting a Yukawa potential $V_q = V_z/(q^2 + \mu^2)$; 
this leads to \begin{eqnarray}
\kappa_z \sim \frac{N c^2 \hbar k_B^4 \mu^4 T^2 T_z}{a n_{\rm imp}
V_z^2} \times \left(\frac{T}{T_z}\right)^{2/\nu - 3}.
\label{kappaz} \end{eqnarray} Here $a$ is the spacing between the layers,
and $T_z$ is the spinon bandwidth in temperature units and is
proportional to the spinon velocity $c$.

\paragraph{(iii.b) Visons.}
The visons are thermally excited across an energy gap, $\Delta_v$,
and so can be considered to be a dilute Boltzmann gas of particles
of mass $m_v$. We assume there are $N_v$ species of visons. The
visons see the background filling of spins as a magnetic flux
through the plaquette on the dual lattice, and hence the dynamics
of visons can be well described by a fully-frustrated quantum
Ising model on the honeycomb lattice. Detailed calculations show
that there are four minima of the vison band with an emergent O(4)
flavor symmetry at low energy ~\cite{sondhi}, therefore $N_v = 4$.
As with the spinons, the visons are assumed to scatter off
impurities of density $n_{\rm imp}$ with, say, a Yukawa potential
$V_q = V_{v}/(q^2 + \mu^2)$. We use the fact that at low $T$, and
for a large vison mass $m_v$, the visons are slowly moving. So
each impurity scattering event can be described by a $T$-matrix $=
[ m_v \ln (1/k)/\pi]^{-1}$ characteristic of low momentum
scattering in two dimensions. Application of Fermi's golden rule
then yields a vison scattering rate $1/\tau_v^{\rm imp} = \pi^2
n_{\rm imp} / (m_v \ln^2 (1/k))$. This formula becomes applicable
when $\ln (1/k) \times V_v/(\hbar^2\mu^2/2m_v) \gg 1$ {\em i.e.\/}
the impurity potential becomes nonperturbative. We can now insert
this scattering rate into a standard Boltzmann equation
computation of the thermal conductivity $\kappa_v = 2 k_B^2 T n_v
\tau_v^{\rm imp}/m_v$, where $n_v$ is the thermally excited vison
density and the typical momentum $k \sim (m_v k_B T)^{1/2}$, to
obtain
\begin{equation}
\kappa_v = \frac{N_v m_v k_B^3 T^2 \ln^2 (T_v/T) e^{-
\Delta_v/(k_B T)}}{4 \pi \hbar^3 n_{\rm imp} a}. \label{kappa}
\end{equation}
Here $T_v$ is some ultraviolet cutoff temperature which can be
taken as the vison bandwidth. Note that for a large density of
states of vison excitations, {\em i.e.\/} a large $m_v$, the
prefactor of the exponential can be large. Similar calculations
will not lead to a logarithmic divergence for the critical spinon
$z$ due to the positive anomalous dimension of $|z|^2$, and
therefore the impurity scattering of spinons is perturbative for
$V_z/(c\mu\hbar)^2 < 1$.

\begin{figure}
\includegraphics[width=1.8in]{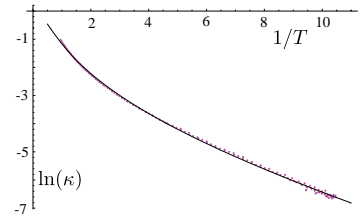}
\caption{Fit of the $T$ dependence of the vison thermal
conductivity in  Eq. (\ref{kappa}) to the thermal conductivity
measurements by Yamashita {\em et al.} \cite{yamashita}; $T_v$,
$\Delta_v$ and the overall prefactor were the fit parameters.}
\label{plot1}
\end{figure}

\begin{figure}
\includegraphics[width=1.8in]{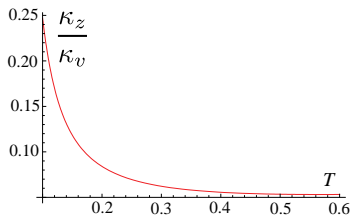}
\caption{Ratio of the thermal conductivity of spinons to visons in
Eq. (\ref{ratioeq})} \label{ratio}
\end{figure}

Using Eq.~(\ref{kappa}), we fit the thermal conductivity measured
by M.~Yamashita {\em et al.\/} in Ref.~\onlinecite{yamashita} by
tuning parameters $T_v$ and $\Delta_v$. The best fit values are
$T_v = 8.15K$, and $\Delta_v \equiv \Delta_\kappa = 0.238K$, as
shown in Fig. \ref{plot1}.
For consistency check, we calculate the ratio between the thermal
conductivities contributed by spinons and visons using Eq.
(\ref{kappaz}) and Eq. (\ref{kappa}) and assuming moderate spinon
impurity strength $V_z/(c\mu\hbar)^2 \sim 1 $: \begin{eqnarray}
\frac{\kappa_z}{\kappa_v} &\sim& \frac{k_B T_z}{m_v c^2}\times
\left(\frac{T}{T_z}\right)^{2/\nu - 3}\frac{1}{(\ln T_v/T)^2}
e^{\Delta_v/k_B T} \cr\cr   &\sim&  \frac{T_v}{T_z}\times
\left(\frac{T}{T_z}\right)^{2/\nu - 3}\frac{1}{(\ln T_v/T)^2}
e^{\Delta_v/(k_B T)}. \label{ratioeq} \end{eqnarray} We plot this ratio in
Fig.~\ref{ratio}, with $T_z \sim J_2 = 250$ K and other parameters
as above, for the experimentally relevant temperature between
$0.1$~K and $0.6$~K; we find consistency because $\kappa$ is
dominated by the vison contribution. The vison dispersion is quadratic
above the vison gap, and this leads to a $T$-independent $\gamma = C_p/T$
when $T > \Delta_v$, as observed in experiments \cite{kanoda2}. Our
estimate of the vison bandwidth, $T_v$, is also consistent with a
peak in both $C_P$ \cite{kanoda2} and $\kappa$ \cite{yamashita} at
a temperature close to $T_v$.

The vison gap, $\Delta_v$,  obtained here is roughly the same as
the temperature at which the $1/T_1$ of NMR starts to deviate from
the low temperature scaling of Eq.~(\ref{Gamma}) \cite{kanoda1}.
When $T$ is above $\Delta_v$,  thermally activated visons will
proliferate. We discuss a theory of the spin dynamics in this
thermal vison regime in the supplement, and find
 a $1/T_1$ with a weaker $T$ dependence
 compared to that present for $T< \Delta_v$. These
observations are qualitatively consistent with the NMR data for
0.25 $<T<$ 10 K \cite{kanoda1}. 

Ref.~\onlinecite{yamashita} also measured the thermal
conductivity, in an applied field $H$ up to 10 T. There was little
change in $\kappa$ for $H< 4 $ T. As $H$ couples to the conserved
total spin, it only appears as an opposite ``chemical potential"
term for $z_\alpha$, modifying the temporal derivative
$(\partial_\tau +  (H/2) \sigma^z) z^\dagger (\partial_\tau -
(H/2) \sigma^z) z$. At the quantum critical point, this term will
induce a condensate of $z$ {\em i.e.} a non-collinear magnetically
ordered state. We do not expect a significant difference in the
thermal conductivity of the gapless spinons versus gapless
spin-waves across this second order transition. We conjecture that
the change at 4 T is associated with a vison condensation
transition to a valence bond solid, as the field scale is or order
the energy scales noted in the previous paragraph. This transition
is possibly connected to the $H$-dependent broadening of the NMR
spectra \cite{kanoda1}.

We have described the properties of a $Z_2$ spin liquid, on the verge of a
transition to an magnetically ordered state,  We have argued that the
quantum critical spinons describe the NMR observations
\cite{kanoda1}, while the visons (with a small energy gap and
bandwidth) dominate the thermal transport \cite{yamashita}. 

We are very grateful to Minoru Yamashita for valuable discussions
of the results of Ref.~\onlinecite{yamashita}, and to the authors
of Ref.~\onlinecite{yamashita} for permission to use their data in
Fig.~\ref{plot1}. We thank K.~Kanoda, S.~Kivelson, and T.~Senthil
for useful discussions. This research was supported by the NSF
under grant DMR-0757145.

\newpage
\cleardoublepage
\setcounter{figure}{0}
\setcounter{equation}{0}
\onecolumngrid
\begin{center}
{\large\bf Dynamics and transport of the $Z_2$ spin liquid: \\
application to $\kappa$-(ET)$_2$Cu$_2$(CN)$_3$}\\
~\\
{\large\bf Supplementary information.}\\
~\\
Yang Qi, Cenke Xu, and Subir Sachdev\\
{\em Department of Physics, Harvard University, Cambridge MA 02138, USA}\\
{\small (Dated: February 3, 2009)}\\
~\\~\\
\end{center}
\twocolumngrid

This supplement presents additional details on computations in the main text.
The large $N$ expansion of the nonlinear $\sigma$ model field
theory of the transition between the $Z_2$ spin liquid and the phase with non-collinear
magnetic order is presented in Section~I.
The thermal conductivity of the spinons is considered in Section~II,
and the NMR relaxation rate at temperatures above the vison gap is 
discussed in Section~III.

\section{I. Large-$N$ expansion of nonlinear $\sigma$ model}

\begin{figure}[htbp]
  \centering
  \includegraphics{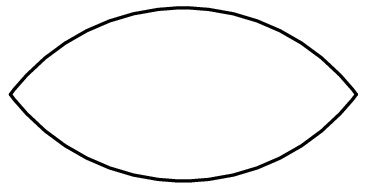}
  \caption{Bubble diagram for the correlation function of $\Pi$. The
    solid line represents propagator given by equation (\ref{eq:G0}).}
  \label{fig:bubble}
\end{figure}

The phase transition between a non-collinear N\'{e}el state and spin
liquid state can be described by the O(4) nonlinear $\sigma$ model as
in Eq.~(1) in the main text
\begin{equation}
  \label{eq:nls-model}
  \mathcal{S}=\frac{1}{g}\int d^2rd\tau [|\partial_\tau z_\alpha|^2+c^2|\nabla_rz_\alpha|^2],
\end{equation}
with the constraint that $|z_1|^2+|z_2|^2=1$. The physical antiferromagnetic order parameter is
related to the O(4) field $z_\alpha$ by the bilinear function \cite{csssup,xucoming}
\begin{equation}
  \label{eq:spin-phys}
  S^i=z_\alpha \sigma_{\alpha\gamma}^y\sigma^i_{\gamma\beta}z_\beta .
\end{equation}
This differs from the collinear case, where the order parameter is linearly
proportional to the field of the O(3) $\sigma$ model.
Therefore, the spin correlation
function is proportional to a bubble diagram of the $z$
field (see Fig. \ref{fig:bubble}) \cite{csssup}:
\begin{equation}
  \label{eq:spin-chi}
  \chi(x, \tau)\sim\Pi(x, \tau)=\left<z_\alpha(x, \tau)z_\beta(x, \tau)
    z_\beta^\ast(0, 0)z_\alpha^\ast(0, 0)\right>.
\end{equation}
The correlation function $\Pi(k, \omega)$ for the O(4) field can then
be calculated using the large-$N$ expansion. The framework of the expansion
can be set up in the disordered phase as follows. First, we rewrite the
constraint as a path integral over a Lagrangian multiplier field
$\lambda$
\begin{equation}
  \label{eq:action-largeN}
  S=\frac{N}{2g}\int d^2rd\tau [|\partial_\tau z_\alpha|^2
  +c^2|\nabla_rz_\alpha|^2+i\lambda(|z_\alpha|^2-1)].
\end{equation}
Here the coupling constant $g$ is rescaled from that in
Eq.~(1) in the main text to show the $N$ dependence in the large-$N$ limit
explicitly. Integrating out the $n$ field in the above action, the path
integral over $\lambda$ becomes
\begin{align}
  \nonumber
  \mathcal{Z}=\int\mathcal{D}\lambda \exp\left[
    -\frac{N}{2}\bigg(\right.&{\rm Tr}\ln(-c^2\nabla^2-\partial_\tau^2+i\lambda)\\
  \label{eq:pi-largeN}
  &\left.\left.
      -\frac{i}{g}\int d\tau d^2x\lambda
  \right)
  \right] .
\end{align}
Therefore, in the $N\rightarrow\infty$ limit, the path integral is
dominated by the contribution from the classical path, along which
$\lambda$ becomes a constant given by the saddle point equation
\begin{equation}
  \label{eq:lambda-eq}
  \frac{1}{\beta}\sum_{\omega_n}
  \int\frac{d^2k}{(2\pi)^2}\frac{1}{\omega_n^2+c^2k^2+m^2}=
  \frac{1}{g},
\end{equation}
where $m^2=i\lambda$, and $m=\Delta_z^{(0)}$ is the spinon gap in the
$N\rightarrow\infty$ limit.

At the $N=\infty$ order, the $\lambda$ field is treated as a constant, and
the theory contains only free $z_\alpha$ field with mass gap
$\Delta_z$. The full large-$N$ expansion is obtained by including
fluctuations of $\lambda$ controlled by the action in Eq.~(\ref{eq:pi-largeN}): 
the $N^{-n}$ order expansion corresponds to a $n$-loop
correction.

The spin correlation function in Eq.~(\ref{eq:spin-chi}) can be
calculated from the $\Pi(\bm{k}, \omega)$ correlation function for the
$z_\alpha$ field using the large-$N$ expansion. At $N=\infty$ order, the
correlation function is given by a bubble diagram of two free
propagators,
\begin{equation}
  \label{eq:chi0-def}
  \Pi_0(k, i\omega_n)=\int\frac{d^2p}{(2\pi)^2}\frac{1}{\beta}\sum_{\nu_n}
  G_0(\bm{p}, i\nu_n)G_0(\bm{p}+\bm{k}, i\nu_n+i\omega_n),
\end{equation}
where $G_0(p, i\omega_n)$ is the free propagator of $z_\alpha$ field
\begin{equation}
  \label{eq:G0}
  G_0(p, i\omega_n)=\frac{1}{c^2p^2+\omega_n^2+m^2}.
\end{equation}

At the $1/N$ order, the contribution from the fluctuation of the $\lambda$ field
needs to be included at one-loop
level. There are two corrections that need to be included for the bubble
diagram: the self-energy correction and the vertex correction.

\begin{figure}[htbp]
  \centering
  \includegraphics{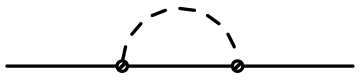}
  \caption{The first term in self-energy correction:
    $\tilde{\Sigma}(k,i\omega_n)$ evaluated in equation
    (\ref{eq:Sigma-tilde}). In this diagram and Fig. \ref{fig:se2} the
    dotted line represents propagator of $\lambda$ field given in
    equation (\ref{eq:pi-largeN}), and interaction vertex between two
    $z_\alpha$ field operators and one $\lambda$ field operator is
    given by the last term in action (\ref{eq:action-largeN}).}
  \label{fig:se1}
\end{figure}

\begin{figure}[htbp]
  \centering
  \includegraphics{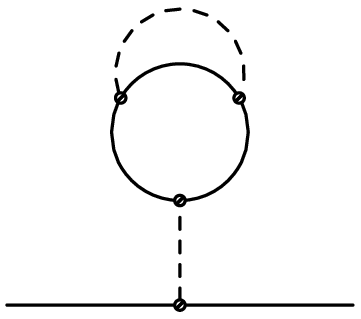}
  \caption{The second term in self-energy correction, which
    corresponds to the last term of equation (\ref{eq:Sigma}). This
    diagram contains two $\lambda$ fields, which scales as $N^{-2}$,
    and also a loop of $z_\alpha$ field, which contributes a factor of
    $N$. Therefore the whole diagram is also at the order of $1/N$.}
  \label{fig:se2}
\end{figure}

\begin{figure}[htbp]
  \centering
  \includegraphics{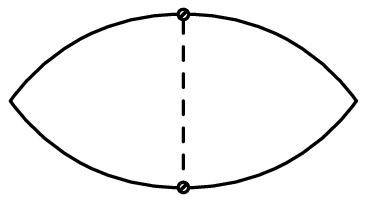}
  \caption{Vertex correction for the bubble diagram, appeared in
    equation (\ref{eq:vertex}).}
  \label{fig:vertex}
\end{figure}

First, the bare propagator in Eq.~(\ref{eq:chi0-def}) needs to be
replaced by a propagator with a self-energy correction at one-loop level\cite{tghs}.
\begin{equation}
  \label{eq:G1}
  G(k, i\omega_n)=\frac{1}{\omega_n^2+c^2k^2+m^2+\Sigma(k, i\omega_n)},
\end{equation}
where the self-energy has two parts. The first part comes from an
insertion of $\lambda$ propagator on $z_\alpha$ propagator shown in
Fig. \ref{fig:se1}:
\begin{widetext}
\begin{equation}
  \label{eq:Sigma-tilde}
  \tilde{\Sigma}(k, i\omega_n)=\frac{2}{N}\frac{1}{\beta}\sum_{\nu_n}
  \int\frac{d^2p}{(2\pi)^2}
  \frac{G_0(\bm{k}+\bm{q}, i\omega_n+i\nu_n)
    -G_0(k, i\omega_n)}{\Pi_0(q, i\nu_n)}.
\end{equation}
The second contribution is given by Fig.~\ref{fig:se2}, and the total
self-energy is
\begin{equation}
  \label{eq:Sigma}
  \Sigma(k, i\omega_n)=\tilde{\Sigma}(k, i\omega_n)
  -\frac{1}{\Pi_0(0, 0)}\frac{1}{\beta}\sum_{\nu_n}
  \int\frac{d^2p}{(2\pi)^2}G_0(p, i\nu)\tilde{\Sigma}(k, i\nu_n)
  G_0(p, i\nu_n) .
\end{equation}

In addition to including the self-energy in the propagators of $\Pi(k,
i\omega_n)$, the vertex correction (see Fig. \ref{fig:vertex}) also needs
to be included.
\begin{equation}
  \label{eq:vertex}
  \Pi^{(1v)}(k, i\omega_n)=
  \frac{2}{N}\frac{1}{\beta^2}\sum_{\nu_n,\epsilon_n}
  \int\frac{d^2pd^2q}{(2\pi)^4}
  \frac{G_0(p, i\nu_n)G_0(\bm{p}+\bm{q}, i\nu_n+i\epsilon_n)
    G_0(\bm{p}+\bm{k}, i\nu_n+i\omega_n)
    G_0(\bm{p}+\bm{q}+\bm{k}, i\nu_n+i\epsilon_n+i\omega_n)}
{\Pi_0(q, i\epsilon_n)}.
\end{equation}
\end{widetext}


\subsection{A. Local susceptibility}

In this section we consider the behavior of the imaginary part of
dynamical susceptibility at the threshold to creating two spinon
excitations. At $N=\infty$ and zero temperature, the
integral in the expression of $\Pi(k, i\omega_n)$ can be evaluated
analytically from Eq.~(\ref{eq:chi0-def}), and the result is
\begin{equation}
  \label{eq:chi0}
  \Pi_0(k,\omega)=\frac{1}{4\pi\sqrt{c^2k^2-\omega^2}}
  \tan^{-1}\left(\frac{\sqrt{c^2k^2-\omega^2}}{2m}\right).
\end{equation}
The real and imaginary part of above equation have the following
asymptotic behavior when $\omega$ is just above the threshold
\begin{equation}
  \label{eq:chi0-real}
  {\rm Re}\, \Pi_0(k,\omega)=\frac{1}{16\pi m}
  \ln\left( \frac{\omega^2-c^2k^2-4m^2}{16m^2}\right),
\end{equation}
and
\begin{equation}
  \label{eq:chi0-imag}
  {\rm Im}\Pi_0(k,\omega)=\frac{{\rm sgn}(\omega)}{8\sqrt{\omega^2-c^2k^2}}
  \theta(\omega-\sqrt{c^2k^2+4m^2}) .
\end{equation}
Naturally, Eqs~(\ref{eq:chi0-real}) and (\ref{eq:chi0-imag}) are connected 
by a Kramers-Kronig relation.

\begin{figure}[htbp]
  \centering
  \includegraphics{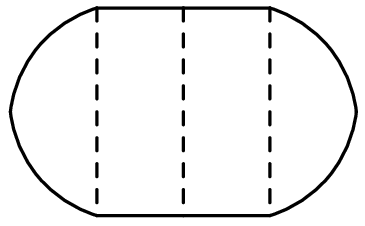}
  \caption{Ladder diagram for vertex correction. A RPA resummation of
    these diagrams is evaluated in equation (\ref{eq:u}).}
  \label{fig:ladder}
\end{figure}

The sharp discontinuity in the imaginary part of susceptibility is an
artifact of the $N=\infty$ limit, and is modified once we add in vertex
corrections. Actually, as the bubble diagram evaluated in equation
(\ref{eq:chi0}) has a logarithmic divergence at the threshold, the
ladder diagrams, which contains all orders of this divergence, should be
summed as in an RPA approximation (see Fig. \ref{fig:ladder}). Since
the bubble is divergent at the threshold, the most divergent
contribution to the ladder diagrams comes from the propagator that is
on shell and at the spinon threshold. When that happens, there is no
momentum transfer through the $\lambda$ propagator. Therefore, we can
approximate the interaction vertex in the RPA summation as the
$\lambda$ propagator at zero momentum:
\begin{equation}
  \label{eq:u}
  u=\frac{2}{N}\frac{1}{\Pi_0(0,0)}=\frac{16\pi}{N}m .
\end{equation}
Therefore the RPA resummation of the ladder diagrams gives
\begin{equation}
  \label{eq:RPA}
  \Pi=\Pi_0+\Pi_0u\Pi_0+\cdots=\frac{\Pi_0}{1-u\Pi_0}.
\end{equation}
Taking the imaginary part of $\chi$, and taking only the most divergent part, we obtain
\begin{equation}
  \label{eq:chi-imag2}
  {\rm Im}\,\Pi=\frac{{\rm Im}\,\Pi_0}{u^2({\rm Re}\, \Pi_0)^2}.
\end{equation}
Pluging into Eqs~(\ref{eq:chi0-real}) and (\ref{eq:chi0-imag}), we obtain
\begin{equation}
  \label{eq:chi-imag3}
  {\rm Im}\,\Pi(k, \omega)=
  \frac{N^2{\rm sgn}(\omega)}
  {8\sqrt{\omega^2-c^2k^2}}
  \frac{\theta\left(|\omega|-\sqrt{c^2k^2+4m^2}\right)}
  {\ln^2\left(\displaystyle \frac{\omega^2-c^2k^2-4m^2}{16m^2}
      \right)}.
\end{equation}

In order to relate this result for $\Pi$ to the physical spin
correlation function $\chi$, we need to insert the proportional
constant in Eq.~(\ref{eq:spin-chi}). Combined with the spectral
weight from higher loop corrections, this gives the constant
$\mathcal{A}$ appearing in Eq.~(2) in the main text. The mass gap
of the spinon $\Delta_z^0=m$ also receives higher loop corrections and
becomes $\Delta_z$ in general. In addition, near the threshold, the
factor of $\sqrt{\omega^2-c^2k^2}$ is approximately
$2\Delta_z$. Therefore the above equation can be rearranged into
\begin{equation}
  \label{eq:imchi-prep}
  {\rm Im}\chi(k, \omega)=\frac{\mathcal{A}N^2{\rm sgn}(\omega)}{16\Delta_z}
  \frac{\theta\left(|\omega|-\sqrt{c^2k^2+4\Delta_z^2}\right)}
  {\ln^2\left(\frac{\omega^2-c^2k^2-4\Delta_z^2}{16\Delta_z^2}\right)}.
\end{equation}
The overall scaling ${\rm Im}\chi\sim\Delta_z^{-1}$ is a result at
$N=\infty$, and shall be refined to
${\rm Im}\chi\sim\Delta_z^{2-\bar{\eta}}$ when higher loop corrections are
included, where $\bar{\eta}$ is the scaling component appearing in
Eq.~(2) in the main text. With this correction, the above equation
becomes
\begin{equation}
  \label{eq:imchi-final}
  {\rm Im}\chi(k, \omega)=\frac{\mathcal{A}N^2{\rm sgn}(\omega)}
  {16\Delta_z^{2-\bar{\eta}}}
  \frac{\theta\left(|\omega|-\sqrt{c^2k^2+4\Delta_z^2}\right)}
  {\ln^2\left(\frac{\omega^2-c^2k^2-4\Delta_z^2}{16\Delta_z^2}\right)},
\end{equation}
which is the same as Eq.~(4) in the main text with
$\mathcal{C}=N^2/16$.

\subsection{B. Relaxation rate at order $1/N$}

In this section we calculate the relaxation rate at $1/N$ order, at
finite temperature above the critical point. We will show that, in the
leading order of $1/N$ expansion, there is a singular term proportional
to ${\ln N}/{N}$.

Following Eq.~(5) in the main text, we calculate
\begin{equation}
  \label{eq:Gamma}
  \Gamma=\lim_{\omega\rightarrow0}\frac{1}{\omega}\int\frac{d^2q}{(2\pi)^2}
  {\rm Im}\chi_L(q, \omega).
\end{equation}

The singularity arises from the $1/N$ self-energy,
by replacing $\chi_L$ with $\Pi$ calculated with the full Green's function
in Eq.~(\ref{eq:G1}).
\begin{equation}
  \label{eq:Pi-G}
  \Pi(q, i\omega_n)=\int\frac{d^2k}{(2\pi)^2}
  \sum_{\nu_n}G(k, i\nu_n)G(\bm{k}+\bm{q}, i\nu_n+i\omega_n).
\end{equation}
In the critical region, temperature is the only energy
scale. Therefore we have set $\beta=1$ in the above equation, and in the remainder
of this subsection.

Pluging Eq.~(\ref{eq:Pi-G}) into (\ref{eq:Gamma}), we obtain
\begin{equation}
  \label{eq:Gamma-Pi-G}
  \Gamma=\lim_{\omega\rightarrow0}\frac{1}{\omega}
  \int\frac{d^2qd^2k}{(2\pi)^4}
  \sum_{\nu_n}
  {\rm Im} G(k, i\nu_n)G(\bm{k}+\bm{q}, i\nu_n+i\omega_n).
\end{equation}
Using the frequency summation identity
\begin{gather}
  \nonumber
  \lim_{\omega\rightarrow0}\frac{1}{\omega}{\rm Im}\sum_{\nu_n}
  G_1(i\nu_n)G_2(i\nu_n+i\omega_n)\\
  \label{eq:fs-id}
  =\int_{-\infty}^\infty\frac{d\epsilon}{2\pi}
  \frac{{\rm Im} G_1(\epsilon) {\rm Im} G_2(\epsilon)}{2\sinh^2
    \frac{\epsilon}{2}},
\end{gather}
and changing the variable in the second integral from $q$ to $\bm{k}+\bm{q}$, we obtain
\begin{equation}
  \label{eq:Gamma-square}
  \Gamma=\int_{-\infty}^\infty\frac{d\epsilon}{2\pi}
  \frac{A(\epsilon)^2}{2\sinh^2
    \frac{\epsilon}{2}},
\end{equation}
where
\begin{equation}
  \label{eq:func-A}
  A(\epsilon)=\int\frac{d^2k}{(2\pi)^2}{\rm Im}\, G(k, \epsilon),
\end{equation}
and the Green's function includes self-energy correction at $1/N$
order
\begin{equation}
  \label{eq:G-1overN}
  G(k, \omega)=\frac{1}{c^2k^2+m^2-\omega^2+\Sigma(k, \omega)}.
\end{equation}

Below we will see that the imaginary part of the
self-energy leads to a ${\ln N}/{N}$ term, which is more divergent than
the $1/N$ correction from the real part. So if we ignore the real part
of the self-energy for the moment, the imaginary part of Green's
function is
\begin{equation}
  \label{eq:imG}
  {\rm Im} \, G(k, \omega)=\frac{{\rm Im}\, \Sigma(k,\omega)}{(c^2k^2+m^2-\omega^2)^2+
    [{\rm Im}\, \Sigma(k,\omega)]^2},
\end{equation}
where $\Sigma(k,\omega)$ is of order $1/N$. For the case that
$k^2+m^2-\omega^2\neq0$, the integrand can be expanded to the order of
$1/N$ by ignoring the ${\rm Im}\Sigma$ term in the denominator. Therefore,
we expand ${\rm Im}\Sigma$ around the quasiparticle pole
$ck_0=\sqrt{\omega^2-m^2}$:
\begin{eqnarray}
   \Sigma(k, \omega)&=&\Sigma(k_0, \omega) \nonumber \\ &+&
  \Sigma^\prime(k_0,\omega)(k^2-k_0^2)+O((k^2-k_0^2)^2) ;
   \label{eq:Sigma-expansion}
\end{eqnarray}
the integral of the second term does not have a singularity because
$k^2-k_0^2$ is an odd function, and higher order terms are also not
singular. Hence these terms result in regular corrections of the order
$1/N$. However, integrating the constant term is divergent near the
pole if the ${\rm Im}\Sigma$ term is ignored. Therefore it needs to be put
back and the most divergent term in $A(\omega)$ is
\begin{equation}
  \label{eq:Aw-sing}
  A(\omega)\sim\int\frac{d^2k}{(2\pi)^2}
  \frac{{\rm Im}\Sigma(k_0,\omega)}{c^2(k-k_0)^2
    +[{\rm Im}\Sigma(k_0,\omega)]^2},
\end{equation}
and the result of this integral is
\begin{equation}
  \label{eq:Aw-atan}
  A(\omega)\sim\frac{1}{8c^2}+\frac{1}{4\pi c^2}\arctan\left[
    \frac{\omega^2-m^2}{{\rm Im}\Sigma(\sqrt{\omega^2-m^2},\omega)}
  \right].
\end{equation}
The function $A(\omega)$ can be expanded to the first two orders of
${\rm Im}\Sigma$ as
\begin{equation}
  \label{eq:Aw-atan2}
  A(\omega)\sim\frac{1}{4c^2}-\frac{1}{4\pi c^2}
    \frac{{\rm Im}\Sigma(\sqrt{\omega^2-m^2},\omega)}{\omega^2-m^2}.
\end{equation}
Pluging this into Eq.~(\ref{eq:Gamma-square}), we obtain
\begin{eqnarray}
  \Gamma &=&\frac{1}{c^2}\int_m^\infty\frac{d\epsilon}{2\pi}
  \frac{1}{\sinh^2(\epsilon/2)} \nonumber \\
  &\times& 
  \left[\frac{1}{16}-\frac{1}{8\pi}
    \frac{{\rm Im}\Sigma(\sqrt{\epsilon^2-m^2},\omega)}{\epsilon^2-m^2}
  \right].  \label{eq:Gamma-calc}
\end{eqnarray}
The first term resembles the relaxation rate in the $N=\infty$ limit, and
the second term yields a ${1}/({N\ln N})$ correction because the
integrand diverges when $\epsilon\rightarrow m$
\begin{equation}
  \label{eq:Gamma-log}
  \Gamma^{(1)}\sim\frac{1}{16\pi^2c^2}
  \frac{{\rm Im}\Sigma(0,m)}{2m\sinh^2(m/2)}
  \ln{\rm Im}\Sigma(0,m).
\end{equation}
Here $m$ is the mass gap of spinon in the critical region at
$\beta=1$. In the $N=\infty$ limit it can be evaluated analytically
\[m=\Theta=2\ln\frac{\sqrt{5}+1}{2}. \]
At order $1/N$ it has been calculated that\cite{csysup}
\[\frac{1}{\tau}=-\frac{{\rm Im}\Sigma(0,m)}{2m}=\frac{0.904}{N}. \]
Thus we obtain
\begin{eqnarray}
   \Gamma^{(1)} &\sim& \frac{0.904}{16\pi^2c^2\sinh^2{\Theta/2}}
  \frac{1}{N}
  \ln N \nonumber \\
  &=&\frac{\sqrt{5}-1}{16\pi c^2}0.931\frac{\ln N}{N}.
 \label{eq:Gamma-log-num}
\end{eqnarray}
This is the ${\ln N}/{N}$ correction in Eq.~(7) in the main text.

\section{II. Spinon thermal conductivity}

The general equation for thermal conductivity at 2+1d CFT was
given in Eq. 8 in the paper: \begin{eqnarray} \kappa_z =
sc^2\tau_{\mathrm{imp}}. \end{eqnarray} The entropy density $s$ is given in
the paper by Eq.~(10). Based on simple scaling arguments, the
leading order scaling behavior of momentum relaxation rate
$1/\tau_{\mathrm{imp}}$ reads \cite{hartnollsup} : \begin{eqnarray}
\frac{1}{\tau_{\mathrm{imp}}} \sim |V_{\mathrm{imp}}|^2
T^{d+1-2/\nu}, \end{eqnarray} with random potential $V(r)$ coupling to
$|z|^2$, and $V_{\mathrm{imp}}$ is defined as
$\overline{V(r)V(r^\prime)} = V^2_{\mathrm{imp}}\delta^2(\vec{r} -
\vec{r}^\prime)$. For a randomly distributed impurity with Yukawa
potential $V_q = V_z/(q^2+ \mu^2)$ and density $n_{\mathrm{imp}}$,
we can identify $V^2_{\mathrm{imp}} \sim
n_{\mathrm{imp}}V^2_z/\mu^4$. Compensating the dimension by
inserting the spinon bandwidth $T_z$ and other physical constants,
we obtain the equation for the thermal conductivity of spinons
(Eq. 11 in the paper): \begin{eqnarray} \kappa_z \sim \frac{N c^2 \hbar k_B^4
\mu^4 T^2 T_z}{a n_{\rm imp} V_z^2} \times
\left(\frac{T}{T_z}\right)^{2/\nu - 3}.\end{eqnarray} Notice that this
equation is only applicable to the case with
$1/\tau_{\mathrm{\mathrm{imp}}} \ll T$.

\section{III. Thermal proliferation of visons}

In this section we discuss the regime $T> \Delta_v$, where the visons have thermally
proliferated. As noted in the paper, at these temperatures the $1/T_1$ NMR relaxation rate
is observed to have a plateau \cite{kanoda1sup}. We believe this is a general feature of a
dense vison regime: the presence of visons makes it harder for the spinons to propagate
independently, and so a vector spin model (which has a $T$-independent NMR relaxation rate \cite{csysup})
becomes more appropriate.

Here we will illustrate this qualitative idea in a specific model. Rather than thinking about this
as high $T$ regime for visons, imagine we reach this regime by sending $\Delta_v \rightarrow 0$
at fixed $T$. In other words, we are in the quantum critical region of a critical point where the vison
gap vanishes leading to phase with the visons condensed. We have already argued in the paper that
the spinons are also in the quantum critical region of a transition where the spinons condense. Thus
a description of the spin dynamics in the regime $T> \Delta_v$ is provided by the quantum criticality
of a multicritical point where both the spinons and visons condense. A general theory of such
multicritical points has been discussed in a recent work by two of us. \cite{xucoming}
The NMR relaxation is then given by $1/T_1 \sim T^{\eta_{mc}}$, where $\eta_{mc}$ is the
anomalous dimension of the magnetic order parameter at the spinon-vison multicritical point.

Our only present estimates of $\eta_{mc}$ come from the $1/N$ expansion,
and so it is useful to compare estimates of anomalous dimensions in this expansion
at different quantum critical points. For the regime, $T < \Delta_z$, discussed in the main paper,
the NMR relaxation is controlled by the theory in Eq.~(\ref{eq:nls-model}) describing the condensation
of the spinons alone. Here we have $1/T_1 \sim T^{\overline{\eta}}$ where \cite{csssup}
\begin{equation}
\overline{\eta} = 1 + \frac{64}{3 \pi^2 N}.
\end{equation}
In the higher temperature regime, $T > \Delta_z$, we have $1/T_1 \sim T^{\eta_{mc}}$,
and the same $1/N$ expansion for this exponent at the multicritical point where
both spinons and visons condense yields \cite{xucoming}
\begin{equation}
\eta_{mc} = \overline{\eta}  - \frac{256}{3 \pi^2 N} \times \frac{1}{1 + 256 k^2/(\pi^2 N^2)}.
\end{equation}
Here $k$ is the level of the Chern-Simons theory describing the multicritical point.
The large $N$ expansion is performed with $k$ proportaional to $N$, and the physical values are $k=2$ and $N=4$.

The key point is that $\eta_{mc} < \overline{\eta}$. Hence $1/T_1$ will have a weaker dependence on $T$
for $T> \Delta_z$ than for $T < \Delta_z$.

\end{document}